# DC transferred arc thermal plasma assisted growth of nanoparticles with different crystalline phases

Naveen V. Kulkarni [a], Soumen Karmakar [a], Indrani Banerjee [b], R. Pasricha [c], S. N. Sahasrabudhe [d], A. K. Das [d] and S. V. Bhoraskar [a,*]

[a] Department of Physics, University of Pune, Pune 411007, India

[b] Birla Institute of Technology, Mesra, Ranchi 835215, India

[c] National Chemical Laboratory, Homi Bhabha Road, Pune 411008, India

[d] Laser and Plasma Technology Division, Bhabha Atomic Research Centre, Trombay, Mumbai 400085, India

[*] Corresponding author, address: Department of Physics, University of Pune, Ganeshkhind, Pune 411007, India

Phone No. +91-20-25692678 (Ext. 427)

Fax. No. +91-20-25691684

*E-mail address:* svb@physics.unipune.ernet.in (S.V. Bhoraskar)

**Abstract**

The control of the crystalline phases of the nanoparticles grown in a direct-current transferred-arc plasma-assisted reactor is reported. The crystalline phases of the as synthesized nanoparticles are shown to critically depend on the operating gas pressure. The paper reports about the change in the crystalline phases of three distinct compounds namely aluminium oxide ($Al_2O_3$), aluminium nitride (AlN) and iron oxide ($Fe_xO_y$). The major outcome of the present work is that the phases having higher defect densities are more probable to form at the sub-atmospheric operating pressure. The variations in the crystalline structures are discussed on the basis of the equilibrium defect density formed during the homogeneous nucleation. The as synthesized nanoparticles were examined by X-ray diffraction analysis and transmission electron microscopy. In addition, the confirmatory analysis for the crystalline phases of the as synthesized iron oxides was carried out with the help of Mössbauer spectroscopy.

**1. Introduction**

Thermal plasma technology [1, 6] since its first development has proved itself to be one of the well established interdisciplinary science topics. Typical applications of this technology include material processing and metallurgy [7-11]. On account of its high temperature (up to $10^4$ K), high activity, rapid chemical reactivity, short processing time, controlled atmosphere and chemical flexibility; thermal plasma is capable of providing wide avenues of reaction products useful for industrial applications requiring material processing on large scale and therefore thermal plasma has gained enough potential in industries [12] and environmental application [13]. There have been many reports [14-17] where thermal plasma systems have been employed for synthesizing nano materials of variety of oxides and nitrides. Although there are numerous methods [18-27] available for producing nanosized particles, the method that utilizes thermal plasma is suitable in terms of raising the production rate. Thermal plasma is generated using DC voltages by using transferred or non-transferred arc plasma modes [1, 2]. In many technological applications large quantities of nanopowders are required. Few such applications include ultra-fine metal particles contained in solid propellant [28] effective catalysts [29], in dirt repellent surfaces, scratch proof coatings, environmental friendly fuel cells [30], and All these applications require particles or wires ranging in diameters between 10 to 100nm, where the large surface area of nanomaterials is important and therefore methods should be evolved wherein the material surface can be made functional. More over the crystalline phase of nanomaterials plays an important role in applications which are solely governed by large surface effects. For example, nanocrystalline anatase phase of titanium oxide is a much efficient photo catalyst as compared to its rutile phase [31].

    Number of applications of nanoparticles is growing day by day. This requires the control of properties as required in each of these applications. There are numerous methods to control the size, size distribution, shape and the composition of nanoparticles. However, the crystalline purity of nanoparticles remains a tough nut to crack. In October 2007, it has been reported by D.J.Palmer [32] that there is probably no single technique known, which can synthesize nanoparticles with a desired crystalline phase. The present paper is probably

the first of its kind, which has shown a direction to control the crystalline phases of different nanomaterials with a simple monitoring of the experimental parameters.

Use of the thermal plasma for synthesis of nanomaterials of ceramic and magnetic materials is available in literature [14-17, 33-37]. Mourra and Munz [38-39] have synthesized nanosized aluminium nitride using two-stage transferred arc reactor, whereas, the present paper describes a direct-current transferred arc thermal plasma reactor (DC-TATPR), in which the metal vapors are directly made to interact with the reacting gas molecules to yield the nanoparticles with two different crystalline phases at different controlled pressures. This method also facilitates the use of optical spectroscopy [40-41]. Moreover to the best of our knowledge none of the papers, so far, has reported the influence of reactor parameters in controlling the crystalline phases of nanoparticles in a constricted DC-TATPR. The present paper discusses the influence of the ambient pressure of the thermal plasma reactor on the formation of crystalline phase of the nanoparticles. We have also reported the spectroscopic analyses of the processing plasma during the synthesis of nanoparticles in detail in our previous communications [40-41]. In the present paper, X-ray diffraction (XRD) method was used to assess the crystallinity of the as synthesized nanoparticles, whereas the morphology was studied using the transmission electron microscopy (TEM).

The possibility of controlling the crystalline phase of a reaction product during the synthesis of nanoparticles, by a suitable choice of the reactor parameter, is demonstrated in this paper by considering three specific compounds. These are aluminium oxide, aluminium nitride and iron oxide. All these compounds exhibit their own potentials in terms of various applications. Because of the fine particle size, high surface area and catalytic activity of their surfaces, the transition alumina (especially the γ form) finds applications in industries as adsorbents, catalyst or catalyst carriers, coatings and soft abrasives [42-43]. The excellent stoichiometry and stability of $Al_2O_3$ helps to make it an important constituent of many protective oxide scales formed on surface of high temperature alloys. AlN which has a very high theoretical thermal conductivity of 320 W/(mK) [44-45] and a thermal expansion close to that of silicon, make it very useful for electronic packaging applications [46]. Maghemite, γ-Fe2O3, is useful as high-density recording media because of its excellent ferrimagnetic properties [47-49]. The preparation of ultrafine γ-Fe2O3 particles is of great interest also due to potential applications in ferro fluids, bioprocess, magnetic refrigeration, information storage, gas sensors and color imaging devices. In such situation; particle size, shape and surface chemistry play important roles in controlling properties like saturation magnetization, coercivity, remnant field, blocking temperature and super-paramagnetism. Magnetite (Fe3O4), a member of spinel type ferrites, is used for recording material, pigments, electro-photographic developer and many such applications [50-51].

## 2. Experimental setup and methods

### 2.1. Description of the reactor

The transferred arc plasma torch operated melter consisted of a stainless steel double walled cylindrical reactor chamber as shown in the schematic in Fig. 1a, whereas the actual photograph is shown in Fig. 1b. The torch was mounted on the upper flange and had a provision for vertical movement over a length of 300 mm along the axis of the

Fig. 1. Single zone, DC transferred arc torch operated thermal plasma reactor used to synthesize the nanoparticles: (a) the schematic diagram; (b) photograph.

cylindrical chamber. The cathode was made up of 2% thoriated tungsten having 6mm diameter and was mounted on a water cooled copper base. A balance between the desired current density and heat conduction capacity decides the cathode taper. The average power loss to the cathode was about 2%. An auxiliary anode in the form of a nozzle was provided for setting up the pilot arc and helped in the constriction and stabilization of the plasma jet. Argon was allowed to pass through the torch and its aerodynamic effect helped stabilizing the plasma jet through the constricted anode-nozzle wall, thus reducing the heat loss and protecting the plasma torch nozzle from erosion. The reactor chamber (Ø=350mm) with a height of 600mm was covered at both the ends by double walled stainless steel covering flanges. Water cooled anode holder fitted at the bottom flange, was provided with a movable jack

Two view ports with diameter of 150mm each, were arranged in such a way that the arc column can be directly viewed and image of the column could be focused at the input of the spectrometer through an optical fiber for diagnostic studies using emission spectroscopy.

*2.2. Experimental details*

Alumina, aluminum nitride and iron oxide were synthesized by the process of thermal plasma assisted gas-phase condensation involving homogeneous nucleation. The plasma forming gas was argon which was made to flow (5 lpm), in a predetermined manner through a plasma torch operated by DC power (4 kW), at 100 A. In this reactor aluminium and iron served as the anode materials and oxygen and nitrogen were chosen as the reacting gases during the synthesis of nanoparticles of oxides, nitrides of aluminium and iron. The argon plasma from the torch was allowed to impinge on the metal anode and subsequently transferred its heat to the target; resulting in the vaporization of the target material. Thermal plasma was used to melt 99.99% pure aluminium and iron and the vapors were allowed to react with oxygen and nitrogen, the operating pressures of which were varied in the reactor. The synthesis was carried out at two different operating ambient pressures of 500 and 760 Torr (1Torr=133Nm$^{-2}$). This was achieved by evacuating the chamber to a base pressure of 0.1 Torr with the help of an oil rotary vacuum pump and then purging 99.99% pure oxygen and nitrogen in different experimental conditions in the reactor up to the desired operating level. The powder was collected over a hemispherical stainless steel, water-cooled collector housed at a radial distance of 12 cm from the anode substrate after the growth was quenched on account of mid-way collisional heat transfer. This collection criterion was maintained for allowing the resulting particles to fly almost equal distances before their deposition. After each operation the reactor was allowed to cool down to the room temperature and then opened subsequently for powder collection. The powders, obtained from the collector, were finally characterized for their properties. The degree of crystallinity of the samples was investigated with a Philips 1866 x-ray diffractometer (resolution 0.01°) with Cu K$\alpha$ radiation *($\lambda$ = 1.542 Å)*. For TEM analysis, the powders were dispersed in 99.999% pure Thomas Baker made ethyl alcohol (C$_2$H$_5$OH) by mild sonication for about one hour. TEM analysis was carried out by employing a JEOL 1200 EX microscope. Further, the confirmation of the different phases of the as-synthesized iron oxides was investigated using Mössbauer spectroscopy (Astin S-600, USA) at room temperature.

## 3. Results and discussion

The mechanism by which the nano particles are grown in the gas phase condensation process is governed by the phenomenon of nucleation followed by the growth, the rate of which is controlled by gas phase collisions. As a result of this collisional interaction the growth kinetics are largely affected; leading into different crystalline structure of the products. The results presented in this section provide direct evidence to this fact. The inference is based on the X-Ray diffraction (XRD) measurements which have been analyzed to reveal the structural differences of nanoparticles grown under different reactor conditions. The relative line intensities in the standard pattern, obtained from the ASTM, data have been compared with the similar line intensities fir the corresponding spectral lines in the measured pattern. Such a comparison was useful to find the major component of the crystalline phase.

Fig. 2 shows a comparison between XRD patterns obtained for aluminium oxides synthesized at 500 Torr and 760 Torr of ambient pressures of oxygen.

Fig. 2. X-ray diffraction spectra of nanocrystalline $Al_2O_3$ synthesized at (a) 500 Torr; and (b) 760 Torr of operating pressures. Line-spectra show the standard patters.

The diffraction patterns have been compared with the line spectra for the closest matched ASTM data [52]. Based on these positions and the relative line intensities which are typical for a particular crystal structure, the peaks have been indexed with the values of miller indices (hkl) corresponding to the planes. It is seen that the particles synthesized at 500 Torr confirm to γ-phase wherein, that synthesized at 760 Torr exhibit δ-phase of $Al_2O_3$.

Similarly, Fig. 3 shows the comparison between the XRD patterns of the particles of the aluminium nitride synthesized at ambient pressures of 500 Torr and 760 Torr of nitrogen.

Fig. 3. X-ray diffraction spectra of nanocrystalline AlN synthesized at (a) 500 Torr; and (b) 760 Torr of operating pressures. Line-spectra show the standard patters.

A line by line comparison with the ASTM data [53] confirms that the sample synthesized at 500 Torr is closer to the cubic phase of aluminium nitride whereas the one synthesized at 760 Torr shows hexagonal phase of aluminium nitride. Traces appearing from pure aluminium [54] are also observed in both these spectra at 2θ = 38.4°, 44.4°, 65.1° and 78.2° corresponding to the (hkl) values of (111), (200), (220) and (311) planes respectively arising from un-reacted aluminium [40]. However, the absence of XRD peaks corresponding to un-reacted aluminium in the XRD spectra of aluminium oxide (Fig. 2) may be on account of high enthalpy of formation of $Al_2O_3$ (-1656.860 kJ $mol^{-1}$) [55], which does not allow the formation of aluminium clusters during the oxidation of Al. On the contrary, lower heat of nitridation (-317.98 kJ $mol^{-1}$) [55] favors the growth of Al nanoparticles during the nitridation process.

A similar comparison of the XRD patterns of iron oxide synthesized at two different ambient pressures (500 Torr and 760 Torr) is presented in Fig. 4.

Fig. 4. X-ray diffraction spectra of nanocrystalline iron oxides synthesized at (a) 500 Torr; and (b) 760 Torr of operating pressures. Line-spectra show the standard patters.

Here the powder synthesized at low ambient pressure (500 Torr) of oxygen shows the XRD pattern much closer to that of γ- $Fe_2O_3$ [56]; whereas the powder synthesized at ambient pressure of 760 Torr shows the XRD patterns closely resembling that of $Fe_3O_4$ [56]. These are however few peaks appearing from $Fe_2O_3$. This is on account of then fact that the XRD peals for cubic spinel γ- $Fe_2O_3$ and $Fe_3O_4$ phases are similar having small differences in their corresponding 'd' values and lattice constants [57]. In spite of this the dominant phases in these two sets of nanoparticles of Iron oxides (prepared at 500 Torr and 760 Torr) are different [57].

From JCPDS data, it is clear that all the XRD peaks of γ-$Fe_2O_3$ and $Fe_3O_4$ are similar with small difference in the d values. Though, X-ray diffraction is necessary for accessing the crystalline purity, but it is not sufficient to characterize the phase differences between γ-$Fe_2O_3$ and $Fe_3O_4$. Mössbauer spectroscopy then provides supplementary but confirmatory information about the stoichiometric control in such types of samples [58]. Fig. 5 shows the Mössbauer spectra for the two samples of iron oxide prepared in the present experiments.

Fig. 5. Mössbauer spectra of iron oxides synthesized at (a) 500 Torr; and (b) 760 Torr of operating pressure. The spectra were recorded at room temperature.

Fig. 5a shows a singlet set of six magnetic hyperfine lines for the sample prepared at 500 Torr. The spectrum is de-convoluted and is observed to consist of single sextet, which, on combining with XRD spectrum (Fig. 4a) confirms the presence of single phase γ-$Fe_2O_3$. However Fig. 5b, corresponding to the sample synthesized at 760 Torr, shows the corresponding Mössbauer spectrum. De-convolution of this spectrum reveals the presence of two sub-spectra of sextets, thereby indicating the presence of two sites, namely tetrahedral and octahedral sites, present in $Fe_3O_4$ [58].

The results obtained from XRD analyses thus exclusively indicate that the crystalline phases of the products are controlled by the ambient pressure maintained during the synthesis of oxides and nitride. The crystalline phase obtained at higher ambient pressure, in each of these cases, shows a tendency of formation of a stable crystalline phase of the product. This conclusion is based on the fact that δ- $Al_2O_3$ is described as super lattice of the spinel structure with ordered cation vacancies and more stable (ΔH=1675.69 kJ/mol) [55]compared to γ- $Al_2O_3$ (ΔH=1656.86 kJ/mol) [55] which has been described as less ordered defect structure [43]. Hexagonal aluminium nitride is more stable as compared to cubic aluminium nitride and similarly $Fe_3O_4$ has higher stability as compared to γ- $Fe_2O_3$. The less stable phases are known to be the transition phases of these crystalline products. Consequently these transition phases are more reactive and provide larger functional groups at the surfaces. No doubt each of the products synthesized in the present reactor is crystalline in structure; unlike those sometimes obtained by chemical routs [59-62]. It is expected that on account of their crystallinity these nanoparticles will find useful applications in catalysis.

The morphology of the nanoparticles in each of these cases for the products is studied with TEM micrographs are shown in Fig. 6, 7 and 8.

Fig. 6. Transmission electron micrograpgs of the nanocrystalline particles of $Al_2O_3$ synthesized at (a) 500 Torr; and (b) 760 Torr of operating pressures.

Fig. 7. Transmission electron micrograpgs of the nanocrystalline particles of AlN synthesized at (a) 500 Torr; and (b) 760 Torr of operating pressures.

Fig. 8. Transmission electron micrograpgs of the nanocrystalline particles of iron oxides synthesized at (a) 500 Torr; and (b) 760 Torr of operating pressures.

It is interesting to see that the nanoparticles of aluminium oxide exhibit almost spherical shape whereas aluminium nitride has invariably shown one-dimensional structures along with the particles. The aspect ratio (L/D) of these one-dimensional structures is seen to be around 8-10; under both the experimental conditions. Further details have been reported in our earlier communication [40]. The shapes of the nanoparticles of iron oxides are seen to be multifaceted and the particle size distribution is wide; as inferred from each of these TEM micrographs. Wide particle size distribution should be expected because thermal plasma provides a steep temperature gradient at the peripheries where the reaction occurs. Moreover the particles encounter collisions on their path before reaching the collector.

The influences of ambient pressure in controlling the crystalline phase/structure are undoubtedly complex and can best be understood, to a certain extent, by revisiting the phenomenon of nucleation and growth in a gas phase reaction. The metal vapors ejected from the anode are expected to move along the directions governed by diffusional mechanism as shown in Fig. 9a whereas, the concentration of the metal vapors inside the plasma plume is decided by the evaporation rates which in turn are controlled by the plasma current and the zonal temperature. The change in the ambient pressure can, then, control the diffusional motion of the ejecting metal vapors across the vapor fronts. The presence of metal vapors was observed by the characteristic colors of the plasma plume.

Fig. 9b and 9c show the photographs of the plasma jet recorded during the evaporation of aluminium and iron respectively.

Fig. 9. (a) Conceptual schematic of the vaporization of anode materials with the help of plasma plume and their reaction zone with the ambient gas; (b) photograph of a typical plasma plume recorded during the evaporation of aluminium; (c) photograph of a typical plasma plume recorded during the evaporation of iron; and (d) conceptual formation of a typical nuclei of a nanoparticles.

The typical colors as observed near the periphery of the plasma jet were greenish-blue in case of Al rich plasma and reddish-yellow in case of Fe rich plasma.

At the higher ambient pressure the atomic density of the reacting gas increases thereby increasing its collisional frequency with the metals atoms, which causes energy loss. This results in reducing the local temperature at the growing interface of the embryo consisting of the molecular species. These molecular species consist of highly metastable

AlO in case of aluminium oxide, FeO [63] in case of iron oxide and AlN [40] in case of aluminium nitride. On account of lower temperature (T), i.e. larger super cooling ΔT, the embryo will grow into a stable crystalline structure.

On the other hand when the ambient pressure of reacting gas is lowered, the local temperature T of the growing interface of the embryo (typically as shown in Fig. 9d) is higher (because collision frequency is low) and the metastable structure is preferred. Here the super-cooling ΔT is relatively higher. These metastable structures have higher defect densities; and thus low enthalpy of formations indicated in Table 1.

Table 1. Reactor specifications and the results obtained with arc current: 100Amp, arc voltage 40 Volts, arc length 50 mm and plasma forming gas: argon.

The cation vacancies in each of these transition phases are higher which is in co-ordination with the equilibrium defect density observed at a given temperature.

The crystal structure of the product nuclei is thus sensitive to the amount of super cooling available at the growing front and hence both the nucleation rate and the growth rate are strongly temperature dependent. The overall rate of recrystallisation changes with temperature for a fixed holding time.

## 4. Conclusion

In conclusion, the gas phase condensation method, using the single zone reactor system (DC-TATPR), is shown to yield different crystalline structures of the product compound, consisting of nanoparticles ranging in the size between 10-80 nm. Lower ambient pressure of 500 Torr is seen to yield the major component to the corresponding transition phase in each of the compounds whereas the higher ambient pressure of 760 Torr, in each case has resulted into a stable structure of the product nanoparticles. These experiments have thus successfully explored the possibility of controlling the major crystalline phases in the nanoparticles of oxides and nitrides grown by the plasma evaporation, by controlling the reactor parameters.

**Acknowledgements**

Naveen V. Kulkarni and S.V. Bhoraskar would like to thank BRNS, DAE, Government of India for funding the project. S. Karmakar acknowledges BARC-PU Joint Collaborative research program for the financial support.

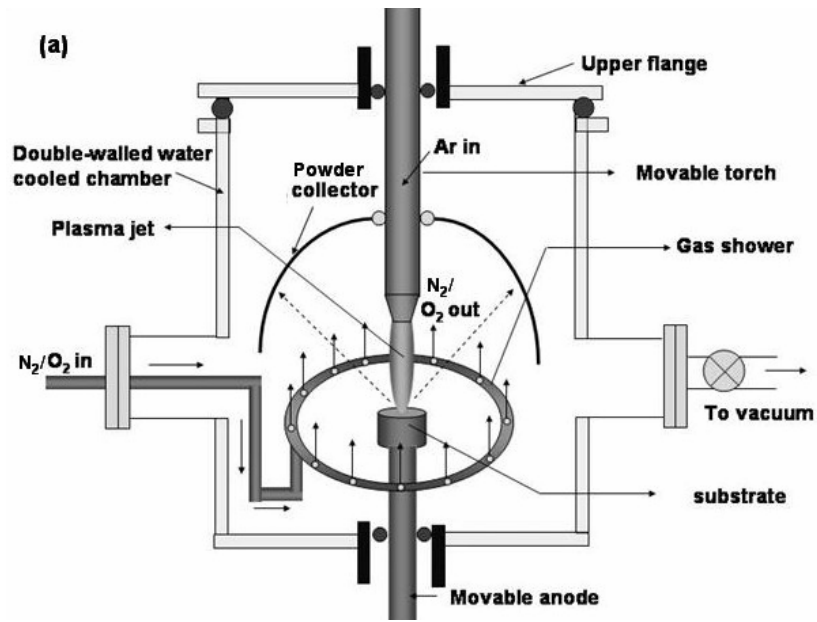

**Fig. 1a.**

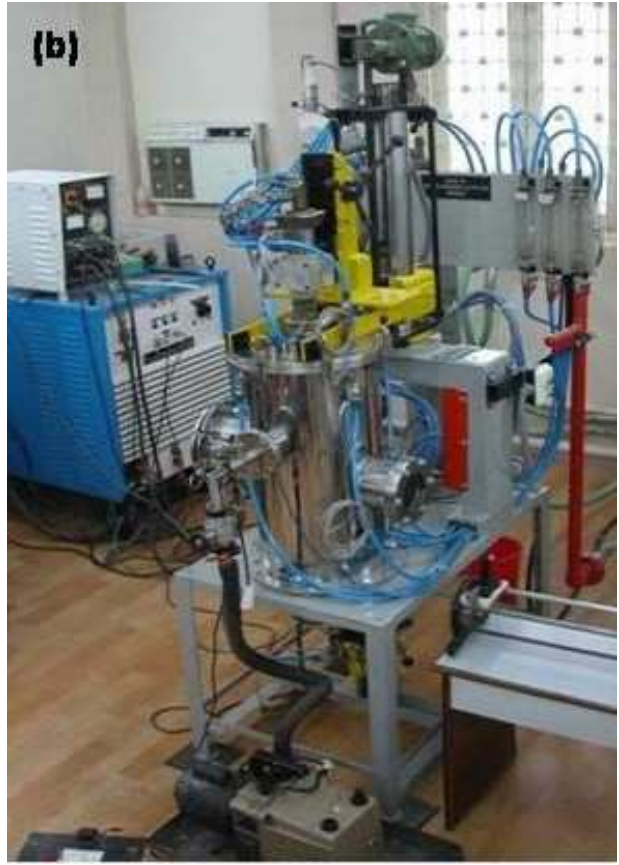

**Fig. 1b.**

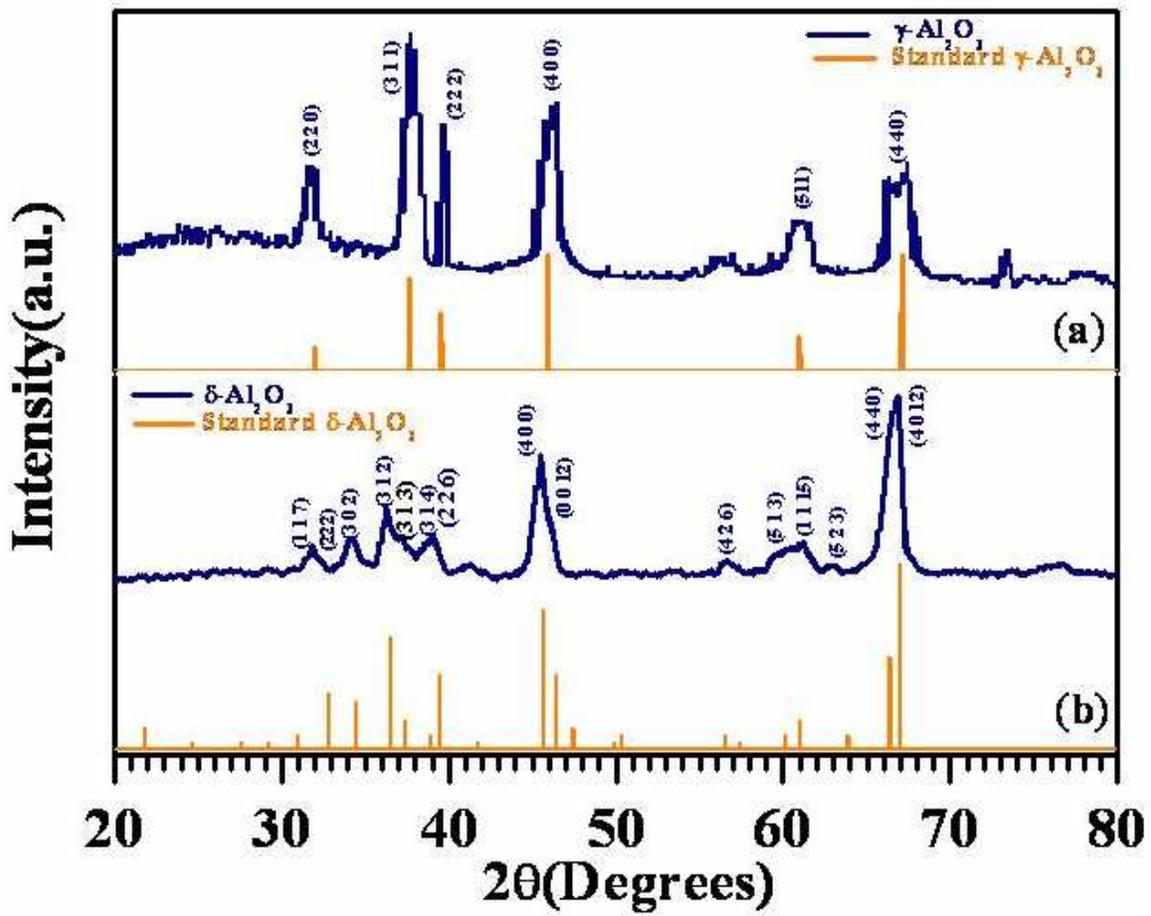

**Fig. 2.**

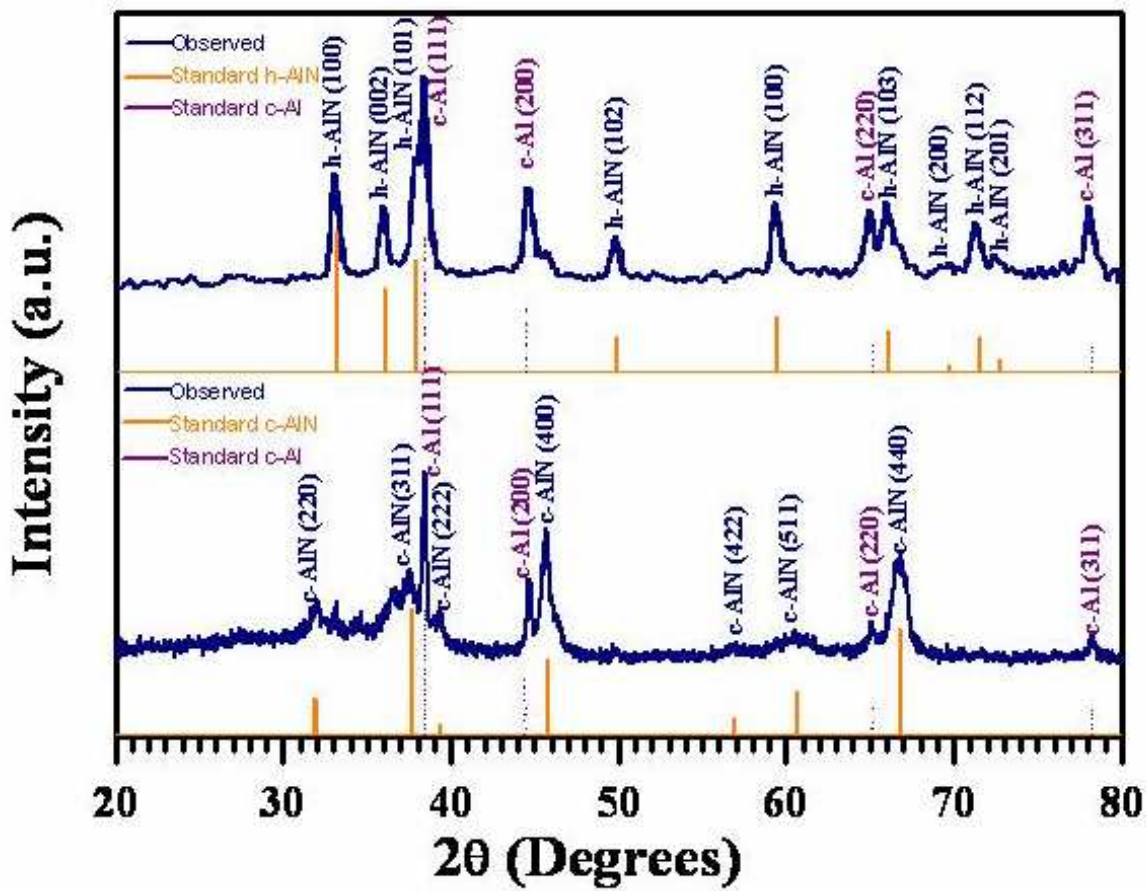

**Fig. 3.**

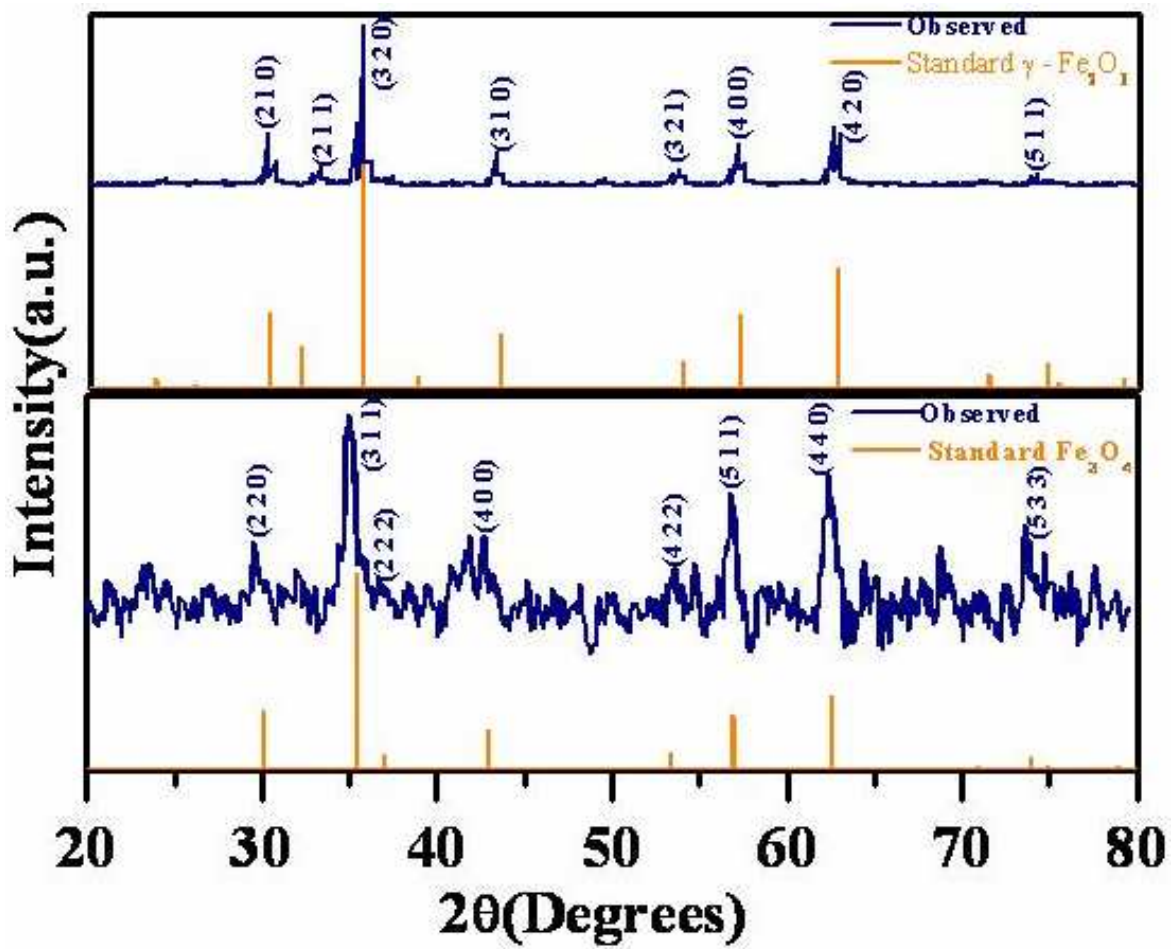

**Fig. 4.**

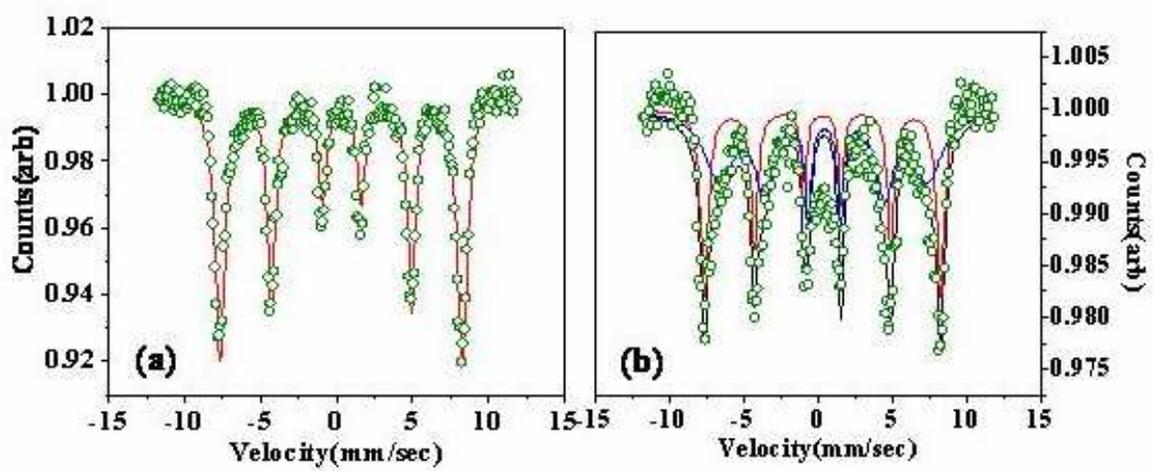

Fig. 5.

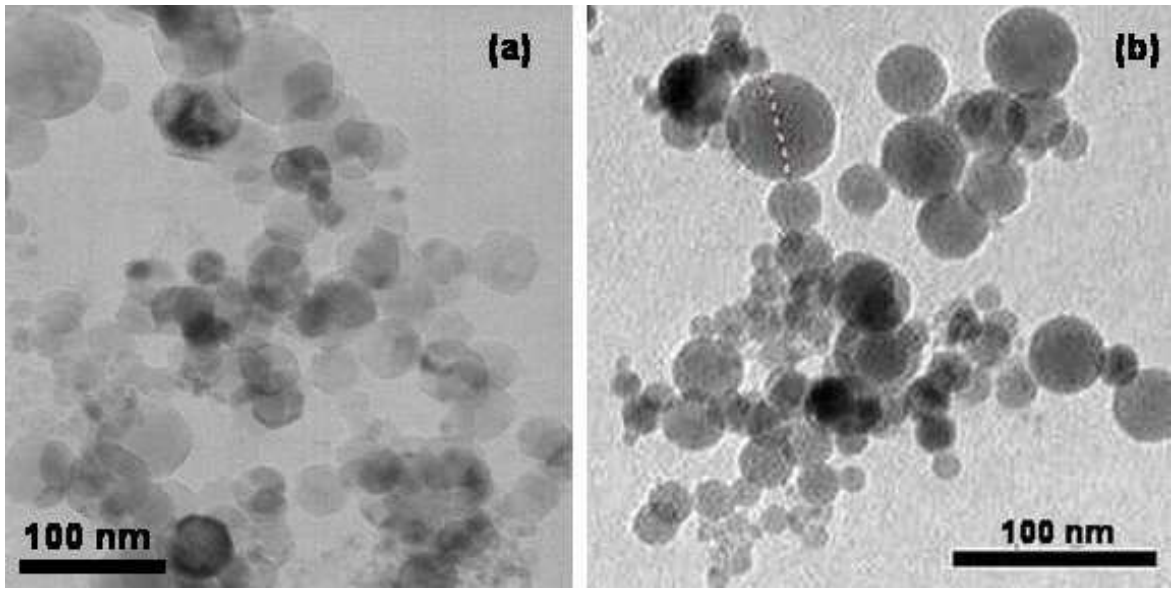

**Fig. 6.**

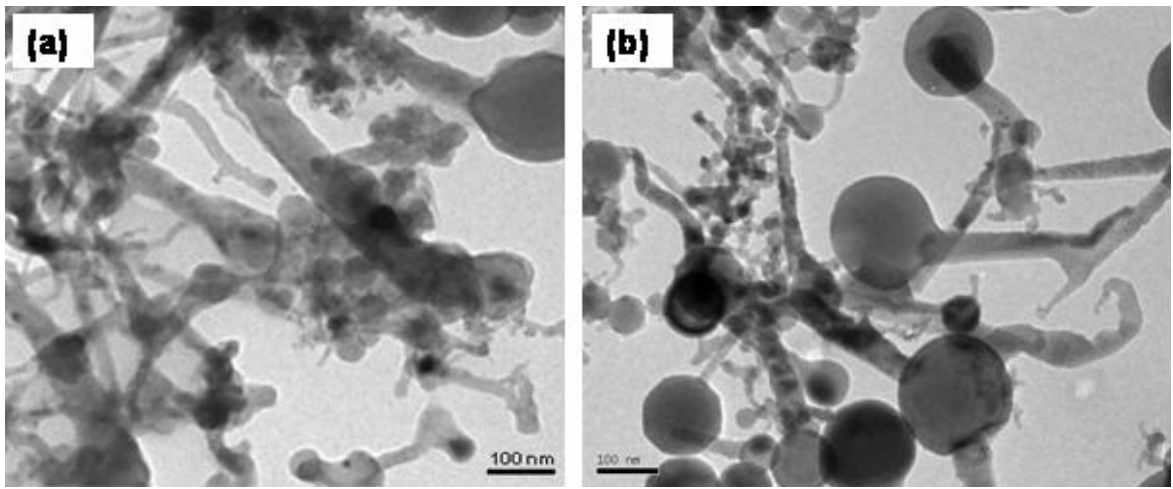

**Fig. 7.**

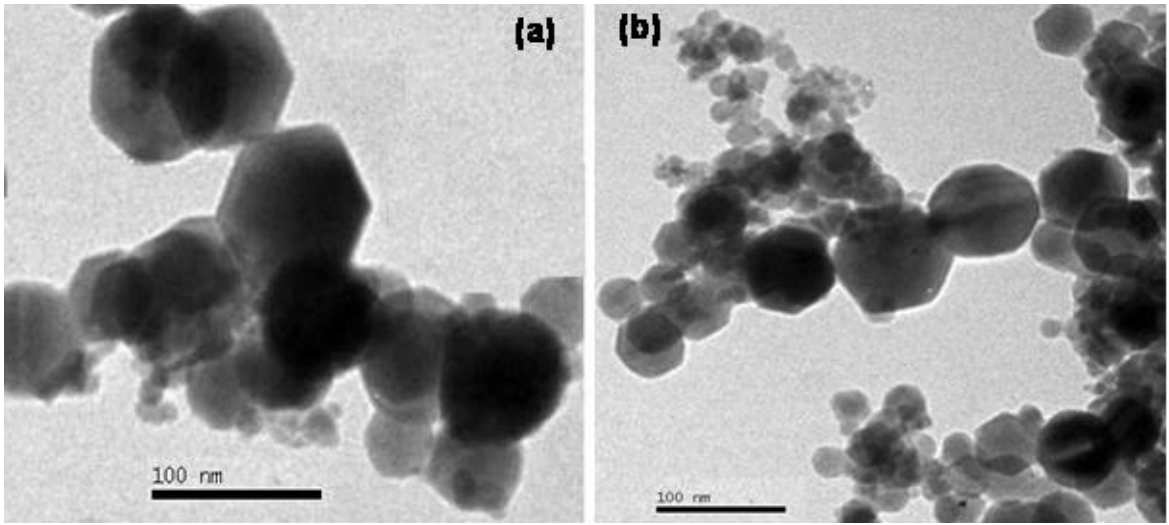

**Fig. 8.**

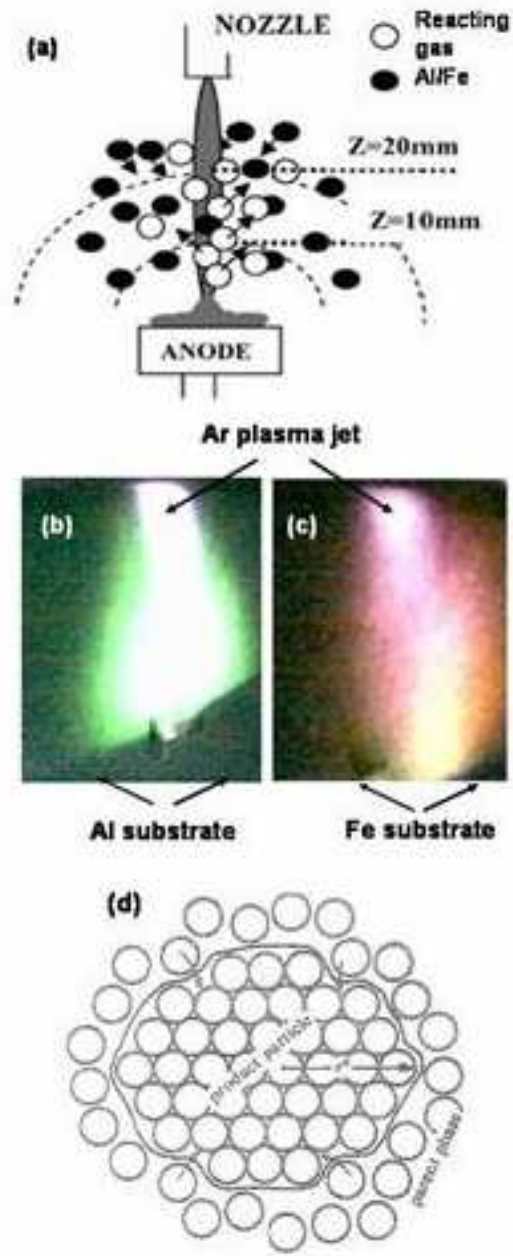

**Fig. 9.**

**Table 1.**

| Anode | Ambient gas | Operating pressure (Torr) | End product | Crystalline phase | Enthalpy of formation (kJ/mol) |
|---|---|---|---|---|---|
| Al | Oxygen | 500 | Aluminium oxide | γ-Al2O3 | -1656.860 |
| Al | Oxygen | 760 | Aluminium oxide | δ-Al2O3 | -1675.690 |
| Al | Nitrogen | 500 | Aluminium nitride | c-AlN | -317.9844 |
| Al | Nitrogen | 760 | Aluminium nitride | h-AlN | -317.9844 |
| Fe | Oxygen | 500 | Iron oxide | $\gamma - Fe_2O_3$ | -823 |
| Fe | Oxygen | 760 | Iron oxide | $Fe_3O_4$ | -1120 |